\documentclass[12pt]{article}

\author{Yu.~M.~Zinoviev
       \thanks{E-mail address: Yurii.Zinoviev@ihep.ru} \\
        {\it Institute for High Energy Physics} \\
        {\it Protvino, Moscow Region, 142280, Russia}}
\title{Toward frame-like gauge invariant formulation\\
for massive mixed symmetry bosonic fields}

\date{}

\begin{document}

\setlength{\unitlength}{1mm}

\maketitle

\begin{abstract}
In this paper, as a first step toward frame-like gauge invariant
formulation for massive mixed symmetry bosonic fields, we consider
mixed tensors, corresponding to Young tableau with two rows with $k
\ge 2$ boxes in the first row and only one box in the second one. We
construct complete Lagrangian and gauge transformations describing
massive particles in (anti) de Sitter space-time with arbitrary
dimension $d \ge 4$ and investigate all possible massless and
partially massless limits.
\end{abstract}

\thispagestyle{empty}
\newpage
\setcounter{page}{1}

\section*{Introduction}

As is well known, in $d=4$ dimensions for the description of arbitrary
spin particles it is enough to consider completely symmetric
(spin-)tensor fields only. At the same time, in dimensions greater
than four in many cases like supergravity theories, superstrings and
high spin theories, one has to deal with mixed symmetry (spin-)tensor
fields \cite{Cur86,AKO86,LM86,Lab89}. There are different approaches
to investigation of such fields both light-cone \cite{Met02,Met04}, as
well as explicitly Lorentz covariant ones (e.g.
\cite{BPT01,BB02,Zin02a,BB06,BKT07,MR07}). For the investigation of
possible interacting theories for high spin particles as well as of
gauge symmetry algebras behind them it is very convenient to use so
called frame-like formulation \cite{Vas80,LV88,Vas88}
(see also \cite{SV06,SV08,Zin08b}) which is a natural generalization
of well known frame formulation of gravity in terms of veilbein
$e_\mu{}^a$ and Lorentz connection $\omega_\mu{}^{ab}$.

Till now, most of the papers on frame-like formulation for mixed
symmetry fields deal with massless case
\cite{Zin03,ASV03,Alk03,ASV05,ASV06,Skv08} (see however
\cite{Zin03a}). The aim of this work is to start an extension of
frame-like formulation to the case of massive mixed symmetry fields.
Namely, we will start a construction of gauge invariant formulation
for such mixed symmetry massive fields in general $(A)dS_d$
space-times with non-zero cosmological constant and arbitrary
space-time dimension $d \ge 4$. There are two general approaches to
gauge invariant description of massive fields. One of them use
powerful BRST approach \cite{BK05,BKRT06,BKL06,BKR07,BKT07,MR07}.
Another one, which we will follow in this work,
\cite{Zin83,KZ97,Zin01,Met06,Zin08b} (see also
\cite{Zin02a,Zin03a,Med03,BHR05,HW05}) is a generalization to
high spins of well known mechanism of spontaneous gauge symmetry
breaking. In this, one starts with appropriate set of massless fields
with all their gauge symmetries and obtain gauge invariant description
of massive field as a smooth deformation.

One of the nice feature of gauge invariant formulation for massive
fields is that it allows us effectively use all known properties of
massless fields serving as building blocks. There are two different
frame-like formulations for massless mixed symmetry bosonic fields.
For simplicity, let us restrict ourselves with mixed symmetry tensors
corresponding to Young tableau with two rows. In what follows we will
denote $Y(k,l)$ a tensor $\Phi^{a_1 \dots a_k,b_1 \dots b_l}$ which is
symmetric both on first $k$ as well as last $l$ indices, completely
traceless on all indices and satisfies a constraint
$\Phi^{(a_1 \dots a_k,b_1) \dots b_l} = 0$, where round brackets mean
symmetrization. In the first approach \cite{ASV03,Alk03,ASV05,ASV06}
for the description of $Y(k,l)$ tensor ($k \ne l$) one use a one-form
$e_\mu{}^{Y(k-1,l)}$ as a main physical field. In this, only one of
two gauge symmetries is realized explicitly and such approach is
very well adapted for the $(A)dS$ spaces. Another formulation
\cite{Skv08} uses two-form $e_{\mu\nu}{}^{Y(k-1,l-1)}$ as a main
physical field in this, both gauge symmetries are realized
explicitly. Such formalism works in flat Minkowski space while
deformations into $(A)dS$ space requires introduction of additional
fields \cite{BMV00}. In this paper we will use the second formalism.
As we have already seen in all cases considered previously and we will
see again in this paper, gauge invariant description of massive fields
always allows smooth deformation into $(A)dS$ space without
introduction of any additional fields besides those that are necessary
in flat Minkowski space so that restriction mentioned above will not
be essential for us.

Mixed symmetry tensor fields have more gauge symmetries compared with
well known case of completely symmetric tensors and, as a result,
gauge invariant formulation for them requires more additional fields
making construction much more involved. In this paper, as a first step
toward gauge invariant frame-like formulation of mixed symmetry
bosonic fields, we consider $Y(k,1)$ tensors for arbitrary $k \ge 2$.
This case turns out to be special and anyway requires separate
consideration. Indeed, in general case $Y(k,l)$, $l > 1$, auxiliary
field analogous to Lorentz connection has to be a two-form
$\omega_{\mu\nu}{}^{Y(k-1,l-1,1)}$, while for the $Y(k,1)$ case one
has to introduce one-form $\omega_\mu{}^{Y(k-1,1,1)}$ instead.
Thus this case turns out to be a natural generalization of simplest
model for $Y(2,1)$ tensor constructed by us before \cite{Zin03a}. The
structure of the paper is simple. In section 1 we reproduce our
results for simplest $Y(2,1)$ tensor. Then, in section 2  we consider
more complex case --- $Y(3,1)$ which shows practically all general
features. At last, in section 3 we construct massive theory for
general $Y(k,1)$ tensor field. In all cases we construct complete
Lagrangian and gauge transformations describing massive particles in
$(A)dS_d$ spaces with arbitrary cosmological constant and arbitrary
space-time dimension $d \ge 4$ and investigate all possible massless
and partially massless limits \cite{DW01,DW01a,DW01c,Zin01,SV06}.

\section{Tensor $Y(2,1)$}

In this case frame-like formulation requires two tensors \cite{Zin03}:
two-form $\Phi_{\mu\nu}{}^a$ as a main physical field and one-form
$\Omega_\mu{}^{abc}$, antisymmetric on $abc$, as analogue of Lorentz
connection. To describe correct number of physical degrees of freedom,
massless Lagrangian has to be invariant under the following gauge
transformations:
\begin{equation}
\delta \Phi_{\mu\nu}{}^a = \partial_{[\mu} \xi_{\nu]}{}^a +
\eta_{\mu\nu}{}^a, \qquad \delta \Omega_\mu{}^{abc} = \partial_\mu
\eta^{abc}
\end{equation}
where $\eta^{abc}$ is completely antisymmetric. Note that
$\xi$-transformations are reducible, i.e.
$$
\xi_\mu{}^a = \partial_\mu \chi^a \qquad \Rightarrow \qquad 
\delta \Phi_{\mu\nu}{}^a = 0
$$
One of the advantages of frame-like formulation is the possibility to
construct an object ("torsion") out of first derivatives of main
physical field $\Phi_{\mu\nu}{}^a$ which is invariant under
$\xi$-transformations:
$$
T_{\mu\nu\alpha}{}^a = \partial_\mu \Phi_{\nu\alpha}{}^a +
\partial_\alpha \Phi_{\mu\nu}{}^a + \partial_\nu \Phi_{\alpha\mu}{}^a
= \partial_{[\mu} \Phi_{\nu\alpha]}
$$
To find a correct form of massless Lagrangian one can use the
following simple trick. Let us consider an expression:
$$
\left\{ \phantom{|}^{\mu\nu\alpha\beta}_{abcd} \right\}
\Omega_\mu{}^{abc} T_{\nu\alpha\beta}{}^d, \qquad
\left\{ \phantom{|}^{\mu\nu\alpha\beta}_{abcd} \right\} =
\delta_a^{[\mu} \delta_b^\mu \delta_c^\alpha \delta_d^{\beta]}
$$
and make a substitution $T_{\mu\nu\alpha}{}^a \rightarrow
\Omega_{[\mu,\nu\alpha]}{}^a$. We obtain:
$$
\left\{ \phantom{|}^{\mu\nu\alpha\beta}_{abcd} \right\}
\Omega_\mu{}^{abc} T_{\nu\alpha\beta}{}^d 
\quad \Rightarrow \quad
\left\{ \phantom{|}^{\mu\nu\alpha\beta}_{abcd} \right\}
\Omega_\mu{}^{abc} \Omega_{\nu,\alpha\beta}{}^d
\quad \Rightarrow \quad
\left\{ \phantom{|}^{\mu\nu}_{ab} \right\}
\Omega_\mu{}^{acd} \Omega_\nu{}^{bcd}
$$
Thus we will look for massless Lagrangian in the form:
$$
{\cal L}_0 = a_1 \left\{ \phantom{|}^{\mu\nu}_{ab} \right\}
\Omega_\mu{}^{acd} \Omega_\nu{}^{bcd} + a_2
\left\{ \phantom{|}^{\mu\nu\alpha\beta}_{abcd} \right\}
\Omega_\mu{}^{abc} T_{\nu\alpha\beta}{}^d 
$$
It is (by construction) invariant under the $\xi$-transformations,
while invariance under $\eta$-transformations requires $a_1 = - 9
a_2$. We choose $a_1 = - 3$, $a_2 = \frac{1}{3}$ and obtain finally:
\begin{equation}
{\cal L}_0 = - 3 \left\{ \phantom{|}^{\mu\nu}_{ab} \right\}
\Omega_\mu{}^{acd} \Omega_\nu{}^{bcd} + 
\left\{ \phantom{|}^{\mu\nu\alpha\beta}_{abcd} \right\}
\Omega_\mu{}^{abc} \partial_\nu \Phi_{\alpha\beta}{}^d 
\end{equation}

All things are very simple in a flat Minkowski space, but if one tries
to consider a deformation of this theory into $(A)dS$ space then it
turns out to be impossible \cite{BMV00}. Thus we turn to the massive
particle and consider the most general case --- massive particle in
$(A)dS$ space with arbitrary cosmological constant. First of all, we
have to determine which additional fields we need to construct gauge
invariant formulation of such massive particle. In general, for each
gauge transformation of main physical field we need appropriate
Goldstone field but in most cases this Goldstone field turns out to be
gauge field by itself so we need Goldstone fields of second order and
so on. But for the mixed symmetry bosonic fields one has to take into
account reducibility of gauge transformations. Let us illustrate this
procedure on our present (simplest) case. Our main physical field
$Y(2,1)$ has two gauge transformations with parameters which are
symmetric $Y(2,0)$ and antisymmetric $Y(1,1)$ tensors correspondingly.
Thus we need two primary Goldstone fields corresponding $Y(2,0)$ and
$Y(1,1)$. Both have their own gauge transformations with vector
parameter $Y(1,0)$, but due to reducibility of gauge transformations
of the main field, we have to introduce one secondary Goldstone field
$Y(1,0)$ only. This field has its own gauge transformation with
parameter $Y(0,0)$, but due to reducibility of gauge transformations
of antisymmetric second rank tensor $Y(1,1)$, the procedure stops
here. It is natural to use frame-like formulation for all fields, so
we introduce four pairs of tensors: ($\Omega_\mu{}^{abc}$, 
$\Phi_{\mu\nu}{}^a$), ($\omega_\mu{}^{ab}$, $h_\mu{}^a$),
($\Omega^{abc}$, $\Phi_{\mu\nu}$) and ($\omega^{ab}$, $h_\mu$). 

We start with the sum of kinetic terms for all fields:
\begin{eqnarray}
{\cal L}_0 &=&  - 3 \left\{ \phantom{|}^{\mu\nu}_{ab} \right\}
\Omega_\mu{}^{acd} \Omega_\nu{}^{bcd} + 
\left\{ \phantom{|}^{\mu\nu\alpha\beta}_{abcd} \right\}
\Omega_\mu{}^{abc} D_\nu \Phi_{\alpha\beta}{}^d + \nonumber \\
 && +\left\{ \phantom{|}^{\mu\nu}_{ab} \right\}
\omega_\mu{}^{ac} \omega_\nu{}^{bc} - 
\left\{ \phantom{|}^{\mu\nu\alpha}_{abc} \right\}
\omega_\mu{}^{ab} D_\nu h_{\alpha}{}^c - \nonumber \\
 && - \Omega_{abc}{}^2 + 
\left\{ \phantom{|}^{\mu\nu\alpha}_{abc} \right\}
\Omega^{abc} D_\mu \Phi_{\nu\alpha} + \omega_{ab}{}^2 - 2
\left\{ \phantom{|}^{\mu\nu}_{ab} \right\} \omega^{ab} D_\mu h_\nu
\end{eqnarray}
as well as appropriate set of initial gauge transformations:
$$
\delta_0 \Phi_{\mu\nu}{}^a = D_{[\mu} \xi_{\nu]}{}^a + 
\eta_{\mu\nu}{}^a, \qquad \delta_0 \Omega_\mu{}^{abc} = D_\mu
\eta^{abc}, \qquad
\delta_0 h_\mu{}^a = D_\mu \zeta^a + \chi_\mu{}^a, 
$$
\begin{equation}
\delta_0 \omega_\mu{}^{ab} = D_\mu \chi^{ab}, \qquad
\delta_0 \Phi_{\mu\nu} = D_{[\mu} \xi_{\nu]}, \qquad 
\delta_0 h_\mu = D_\mu \zeta
\end{equation}
where all partial derivatives are replaced by $(A)dS$ covariant ones.
Here and in what follows, we will use the following convention on
covariant derivatives:
\begin{equation}
[ D_\mu, D_\nu ] \xi^a = - \kappa (e_\mu{}^a \xi_\nu - e_\nu{}^a
\xi_\mu ), \qquad \kappa = \frac{2 \Lambda}{(d-1)(d-2)}
\end{equation}
Note, that due to non-commutativity of covariant derivatives such
Lagrangian is not invariant under the initial gauge transformations:
$$
\delta_0 {\cal L}_0 = \kappa \left\{ \phantom{|}^{\mu\nu}_{ab}
\right\} [ 3 (d-3) ( 2 \Omega_\mu{}^{abc} \xi_\nu{}^c + \eta^{abc}
\Phi_{\mu\nu}{}^c ) - (d-2) ( \omega_\mu{}^{ab} \zeta_\nu - \chi^{ab}
h_{\mu\nu}) ]
$$
so we have to take this non-invariance into account later on. Now to
proceed with the construction of gauge invariant formulation for
massive particle, we have to add to the Lagrangian all possible cross
terms of order $m$ (i.e. with the coefficients having dimension of
mass). Moreover, as our previous experience shows, we need to
introduce cross terms for the nearest neighbours only, i.e. main field
with primary Goldstone fields, primary fields with secondary ones and
so on. For the case at hands all possible such terms could be written
as follows:
\begin{eqnarray}
{\cal L}_1 &=& \left\{ \phantom{|}^{\mu\nu\alpha}_{abc} \right\}
[ a_1 \omega_\mu{}^{ab} \Phi_{\nu\alpha}{}^c + a_2 \Omega_\mu{}^{abc}
\Phi_{\nu\alpha} ] + \left\{ \phantom{|}^{\mu\nu}_{ab} \right\} [ a_3
\Omega_\mu{}^{abc} h_\nu{}^c + a_4 \Omega^{abc} \Phi_{\mu\nu}{}^c ] +
\nonumber \\
 && + \left\{ \phantom{|}^{\mu\nu}_{ab} \right\} [ a_5
\omega_\mu{}^{ab} h_\nu + a_6 \omega^{ab} \Phi_{\mu\nu} ] + a_7
\left\{ \phantom{|}^{\mu}_{a} \right\} \omega^{ab} h_\nu{}^b
\end{eqnarray}
Non-invariance of these terms under the initial gauge transformations
could be compensated by the following corrections to gauge
transformations:
\begin{eqnarray}
\delta_1 \Phi_{\mu\nu}{}^a &=& \frac{\beta_1}{12(d-3)} e_{[\mu}{}^a
\zeta_{\nu]} - \frac{3\alpha_1}{(d-3)} e_{[\mu}{}^a \xi_{\nu]}, \qquad
\delta_1 \Omega_\mu{}^{abc} = \frac{\beta_1}{6(d-3)} e_\mu{}^{[a}
\chi^{bc]} \nonumber \\
\delta_1 h_\mu{}^a &=& \beta_1 \xi_\mu{}^a + \frac{4 \rho_0}{d-2}
e_\mu{}^a \zeta, \qquad \delta_1 \omega_\mu{}^{ab} = - 
\frac{\beta_1}{2} \eta_\mu{}^{ab} \nonumber \\
\delta_1 \Phi_{\mu\nu} &=& \alpha_1 \xi_{[\mu,\nu]}, \qquad 
\delta_1 \Omega^{abc} = - 3 \alpha_1 \eta^{abc}, \\
\delta_1 h_\mu &=& \rho_0 \zeta_\mu + \beta_0 \xi_\mu, \qquad
\delta_1 \omega^{ab} = - 2 \rho_0 \chi^{ab} \nonumber
\end{eqnarray}
provided:
$$
a_1 = a_3 = \frac{\beta_1}{2}, \qquad 
a_2 = a_4 = - 3 \alpha_1, \qquad
a_5 = a_7 = 4 \rho_0, \qquad
a_6 = \beta_0
$$
Thus we have $\delta_0 {\cal L}_1 + \delta_1 {\cal L}_0 = 0$ and this
leaves us with variations of order $m^2$ (taking into account
non-invariance of kinetic terms due to non-commutativity of covariant
derivatives) $\delta_0 {\cal L}_0 + \delta_1 {\cal L}_1$. In general,
to compensate this non-invariance one has to introduce mass-like terms
into the Lagrangian as well as appropriate corrections for gauge
transformations. But in this case there are no possible mass-like
terms (the only possible term $\left\{ \phantom{|}^{\mu\nu}_{ab}
\right\} h_\mu{}^a h_\nu{}^b$ is forbidden by $\zeta$-invariance).
Nevertheless, it turns out to be possible to achieve complete
invariance without any explicit mass-like terms just by adjusting the
values for our four main parameters $\alpha_1$, $\beta_1$, $\beta_0$
and $\rho_0$. We obtain:
$$
\rho_0 = \sqrt{\frac{3(d-2)}{4(d-3)}} \alpha_1, \qquad
\beta_0 = - \sqrt{\frac{3(d-2)}{4(d-3)}} \beta_1, \qquad
\beta_1{}^2 - 36 \alpha_1{}^2 = - 12 \kappa (d-3)
$$

Now we are ready to analyze results obtained. First of all, recall
that there is no strict definition of what is mass in $(A)dS$ space
(see e.g. discussion in \cite{Gar03}). Working with gauge invariant
description of massive particles it is natural to define massless
limit as a limit where all Goldstone fields decouple from the main
gauge field. To make analyze more transparent, let us give here a
Figure 1 showing the roles played by our four parameters.
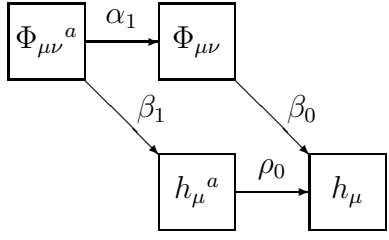
\begin{figure}[htb]
\begin{center}
\begin{picture}(70,32)
\put(10,21){\framebox(10,10)[]{$\Phi_{\mu\nu}{}^a$}}
\put(30,21){\framebox(10,10)[]{$\Phi_{\mu\nu}$}}
\put(30,1){\framebox(10,10)[]{$h_\mu{}^a$}}
\put(50,1){\framebox(10,10)[]{$h_\mu$}}
\put(20,26){\vector(1,0){10}}
\put(22,26){\makebox(6,6)[]{$\alpha_1$}}
\put(20,21){\vector(1,-1){10}}
\put(26,14){\makebox(6,6)[]{$\beta_1$}}
\put(40,21){\vector(1,-1){10}}
\put(46,14){\makebox(6,6)[]{$\beta_0$}}
\put(40,6){\vector(1,0){10}}
\put(42,6){\makebox(6,6)[]{$\rho_0$}}
\end{picture}
\end{center}
\caption{General massive theory for $Y(2,1)$ tensor}
\end{figure}
One can easily see that massless limit is a limit where both $\alpha_1
\rightarrow 0$ and $\beta_1 \rightarrow 0$ simultaneously. But from
the last relation above it is immediately follows that such a limit is
possible in flat Minkowski space ($\kappa = 0$) only. For non-zero
values of cosmological constant one can obtain partially massless
limits instead. Indeed, in $AdS$ space ($\kappa < 0$) one can put
$\alpha_1 = 0$ (and this gives $\rho_0 = 0$). Then our system
decomposes into two disconnected subsystems. One of them with the
fields $\Phi_{\mu\nu}{}^a$ and $h_\mu{}^a$ describe partially massless
theory \cite{BMV00} with the Lagrangian:
\begin{equation}
{\cal L} = {\cal L}_0 (\Phi_{\mu\nu}{}^a) + {\cal L}_0 (h_\mu{}^a) +
\frac{\beta_1}{2} \left\{ \phantom{|}^{\mu\nu\alpha}_{abc} \right\}
\omega_\mu{}^{ab} \Phi_{\nu\alpha}{}^c + 
\frac{\beta_1}{2} \left\{ \phantom{|}^{\mu\nu}_{ab} \right\}
\Omega_\mu{}^{abc} h_\nu{}^c
\end{equation}
which is invariant under the following gauge transformations:
\begin{eqnarray}
\delta \Phi_{\mu\nu}{}^a &=& D_{[\mu} \xi_{\nu]}{}^a + 
\eta_{\mu\nu}{}^a + \frac{\beta_1}{12(d-3)} e_{[\mu}{}^a
\zeta_{\nu]}, \qquad
\delta \Omega_\mu{}^{abc} = D_\mu \eta^{abc} + 
\frac{\beta_1}{6(d-3)} e_\mu{}^{[a} \chi^{bc]} \nonumber \\
\delta h_\mu{}^a &=& D_\mu \zeta^a + \chi_\mu{}^a + 
\beta_1 \xi_\mu{}^a, \qquad \delta \omega_\mu{}^{ab} = D_\mu \chi^{ab}
 - \frac{\beta_1}{2} \eta_\mu{}^{ab}
\end{eqnarray}
where $\beta_1{}^2 = - 12 \kappa (d-3)$. In this, two other fields
$\Phi_{\mu\nu}$ and $h_\mu$ provide gauge invariant description of
massive antisymmetric second rank tensor field. In turn, in $dS$ space
($\kappa > 0$) one can put $\beta_1 = 0$ (and this gives $\beta_0 =
0$). In this case our system also decompose into two disconnected
subsystems. One of them with the fields $\Phi_{\mu\nu}{}^a$ and
$\Phi_{\mu\nu}$ gives another example of partially massless theory
with the Lagrangian:
\begin{equation}
{\cal L} = {\cal L}_0 (\Phi_{\mu\nu}{}^a) + {\cal L}_0 (\Phi_{\mu\nu})
- 3 \alpha_1 \left\{ \phantom{|}^{\mu\nu\alpha}_{abc} \right\}
\Omega_\mu{}^{abc} \Phi_{\nu\alpha} - 3 \alpha_1
\left\{ \phantom{|}^{\mu\nu}_{ab} \right\}
\Omega^{abc} \Phi_{\mu\nu}{}^c 
\end{equation}
which is invariant under the following gauge transformations:
\begin{eqnarray}
\delta \Phi_{\mu\nu}{}^a &=& D_{[\mu} \xi_{\nu]}{}^a + 
\eta_{\mu\nu}{}^a  - \frac{3\alpha_1}{(d-3)} e_{[\mu}{}^a
\xi_{\nu]}, \qquad
\delta \Omega_\mu{}^{abc} = D_\mu \eta^{abc} \nonumber \\
\delta \Phi_{\mu\nu} &=&D_{[\mu} \xi_{\nu]} + \alpha_1
\xi_{[\mu,\nu]}, \qquad 
\delta \Omega^{abc} = - 3 \alpha_1 \eta^{abc}
\end{eqnarray}
where $3 \alpha_1{}^2 = \kappa (d-3)$. In this, two other fields
$h_\mu{}^a$ and $h_\mu$ provide gauge invariant description of
partially massless spin 2 particle \cite{Zin03a,SV06,Zin08b}.

\section{Tensor $Y(3,1)$}

For the description of massless field we will use natural
generalization of simplest example given above. Namely, we introduce
two-form $\Phi_{\mu\nu}{}^{ab}$ which is symmetric and traceless on
$ab$ as a main physical field as well as auxiliary one-form
$\Omega_\mu{}^{abc,d}$ which is completely antisymmetric on $abc$,
traceless and satisfies a constraint $\Omega_\mu{}^{[abc,d]} = 0$. To
provide correct number of physical degrees of freedom massless
Lagrangian has to be invariant under the following gauge
transformations:
\begin{equation}
\delta \Phi_{\mu\nu}{}^{ab} = \partial_{[\mu} \xi_{\nu]}{}^{ab} +
\eta_{\mu\nu}{}^{(a,b)}, \qquad
\delta \Omega_\mu{}^{abc,d} = \partial_\mu \eta^{abc,d}
\end{equation}
Here $\xi_\mu{}^{ab}$ is symmetric and traceless on $ab$, while
$\eta^{abc,d}$ has the same properties on local indices as
$\Omega_\mu{}^{abc,d}$. Note, that these gauge transformations are
also reducible:
$$
\xi_\mu{}^{ab} = \partial_\mu \chi^{ab} \quad \Rightarrow \quad \delta
\Phi_{\mu\nu}{}^{ab} = 0
$$
To construct appropriate massless Lagrangian we will use the same
trick as before. We introduce a "torsion" tensor 
$T_{\mu\nu\alpha}{}^{ab} = \partial_{[\mu} \Phi_{\nu\alpha]}{}^{ab}$
which is invariant under $\xi$-transformations, consider an expression
$\left\{ \phantom{|}^{\mu\nu\alpha\beta}_{abcd} \right\}
\Omega_\mu{}^{abc,e} T_{\nu\alpha\beta}{}^{de}$ and make a
substitution $T_{\mu\nu\alpha}{}^{ab} \Rightarrow 
\Omega_{[\mu,\nu\alpha]}{}^{(a,b)}$. We obtain:
$$
\left\{ \phantom{|}^{\mu\nu\alpha\beta}_{abcd} \right\}
\Omega_\mu{}^{abc,e} T_{\nu\alpha\beta}{}^{de} \quad
\Rightarrow \quad
\left\{ \phantom{|}^{\mu\nu\alpha\beta}_{abcd} \right\}
\Omega_\mu{}^{abc,e} ( \Omega_{\nu,\alpha\beta}{}^{d,e} + 
\Omega_{\nu,\alpha\beta}{}^{e,d} )  \quad \Rightarrow
$$
$$
\left\{ \phantom{|}^{\mu\nu}_{ab} \right\}  [
3 \Omega_\mu{}^{acd,e} \Omega_\nu{}^{bcd,e} +
\Omega_\mu{}^{cde,a} \Omega_\nu{}^{cde,b} ]
$$
Thus we will look for the massless Lagrangian in the form:
$$
{\cal L}_0 = a_ 1 \left\{ \phantom{|}^{\mu\nu}_{ab} \right\}  [
3 \Omega_\mu{}^{acd,e} \Omega_\nu{}^{bcd,e} +
\Omega_\mu{}^{cde,a} \Omega_\nu{}^{cde,b} ] + a_2
\left\{ \phantom{|}^{\mu\nu\alpha\beta}_{abcd} \right\}
\Omega_\mu{}^{abc,e} T_{\nu\alpha\beta}{}^{de}
$$
This Lagrangian is (by construction) invariant under
$\xi$-transformations, while invariance under $\eta$-transformations
requires $a_1 = - 3 a_2$. We choose $a_1 = 1$, $a_2 = - \frac{1}{3}$
and finally obtain:
\begin{equation}
{\cal L}_0 (\Phi_{\mu\nu}{}^{ab})
 = \left\{ \phantom{|}^{\mu\nu}_{ab} \right\}  [
3 \Omega_\mu{}^{acd,e} \Omega_\nu{}^{bcd,e} +
\Omega_\mu{}^{cde,a} \Omega_\nu{}^{cde,b} ] - 
\left\{ \phantom{|}^{\mu\nu\alpha\beta}_{abcd} \right\}
\Omega_\mu{}^{abc,e} \partial_\nu \Phi_{\alpha\beta}{}^{de}
\end{equation}

As in the previous case, it is not possible to deform this massless
Lagrangian into $(A)dS$ space without introduction of additional
fields. Thus we turn to the general case --- massive particle in
$(A)dS$ space with arbitrary cosmological constant. First of all, we
have to determine the set of additional fields which is necessary for
gauge invariant description of such massive particle. Our main gauge
field $Y(3,1)$ has two gauge transformations (combined into one
$\xi_\nu{}^{ab}$ transformation in frame-like formalism) with
parameters corresponding to $Y(2,1)$ and $Y(3,0)$. Recall that these
transformations are reducible with the reducibility parameter
$Y(2,0)$. Thus we have to introduce two primary Goldstone fields ---
$Y(2,1)$ and $Y(3,0)$. The first one also has two gauge
transformations with parameters $Y(1,1)$ and $Y(2,0)$ with the
reducibility $Y(1,0)$, while the second field has one gauge
transformation $Y(2,0)$ only. Taking into account reducibility of main
field gauge transformations it is enough to introduce two secondary
fields $Y(1,1)$ and $Y(2,0)$ only. Both have gauge transformations
with parameters $Y(1,0)$, but due to reducibility of gauge
transformations for $Y(2,1)$ field it is enough to introduce one
additional field $Y(1,0)$. It has its own gauge transformation
$Y(0,0)$, but due to reducibility of gauge transformations for
$Y(1,1)$ field, the procedure stops here. Thus we need six fields ---
$Y(l,1)$, $Y(l,0)$, $1 \le l \le 3$. 

Again we will use frame-like formalism for the description of all
fields and introduce six pairs: ($\Omega_\mu{}^{abc,d}$,
$\Phi_{\mu\nu}{}^{ab}$), ($\omega_\mu{}^{a,bc}$, $h_\mu{}^{ab}$),
($\Omega_\mu{}^{abc}$, $\Phi_{\mu\nu}{}^a$),
($\omega_\mu{}^{ab}$, $h_\mu{}^a$), ($\Omega^{abc}$, $\Phi_{\mu\nu}$)
and ($\omega^{ab}$, $h_\mu$). Note, that here and in what follows we
use the same conventions for the frame-like formulation of $Y(k,0)$
fields as in \cite{Zin08b}. We start with the sum of kinetic terms for
all six fields:
\begin{eqnarray}
{\cal L}_0 &=& \left\{ \phantom{|}^{\mu\nu}_{ab} \right\}  [
3 \Omega_\mu{}^{acd,e} \Omega_\nu{}^{bcd,e} +
\Omega_\mu{}^{cde,a} \Omega_\nu{}^{cde,b} ] - 
\left\{ \phantom{|}^{\mu\nu\alpha\beta}_{abcd} \right\}
\Omega_\mu{}^{abc,e} D_\nu \Phi_{\alpha\beta}{}^{de} - \nonumber \\
 &&  - \left\{ \phantom{|}^{\mu\nu}_{ab} \right\} [ \frac{1}{2}
\omega_\mu{}^{a,cd} \omega_\nu{}^{b,cd} + 
\omega_\mu{}^{c,ad} \omega_\nu{}^{c,bd} ] + 2
\left\{ \phantom{|}^{\mu\nu\alpha}_{abc} \right\}
 \omega_\mu{}^{a,bd} D_\nu h_{\alpha}{}^{cd} - \nonumber \\
 && - 3  \left\{ \phantom{|}^{\mu\nu}_{ab} \right\}
\Omega_\mu{}^{acd} \Omega_\nu{}^{bcd} + 
\left\{ \phantom{|}^{\mu\nu\alpha\beta}_{abcd} \right\}
\Omega_\mu{}^{abc} D_\nu \Phi_{\alpha\beta}{}^d + \nonumber \\
 && + \left\{ \phantom{|}^{\mu\nu}_{ab} \right\}
\omega_\mu{}^{ac} \omega_\nu{}^{bc} - 
\left\{ \phantom{|}^{\mu\nu\alpha}_{abc} \right\}
\omega_\mu{}^{ab} D_\nu h_{\alpha}{}^c - \nonumber \\
 && - \Omega_{abc}{}^2 + \left\{ \phantom{|}^{\mu\nu\alpha}_{abc}
\right\} \Omega^{abc} \partial_\mu \Phi_{\nu\alpha} + \frac{1}{2}
\omega_{ab}{}^2 - \left\{ \phantom{|}^{\mu\nu}_{ab} \right\}
\omega^{ab} \partial_\mu h_\nu
\end{eqnarray}
as well as with appropriate set of initial gauge transformations:
\begin{eqnarray}
\delta_0 \Phi_{\mu\nu}{}^{ab} = D_{[\mu} \xi_{\nu]}{}^{ab} +
\eta_{\mu\nu}{}^{(a,b)}, & \qquad 
&\delta_0 \Omega_\mu{}^{abc,d} = D_\mu \eta^{abc,d} \nonumber \\
\delta_0 h_\mu{}^{ab} = D_\mu \zeta^{ab} + \chi_\mu{}^{ab},
&\qquad &\delta_0 \omega_\mu{}^{a.bc} = D_\mu \chi^{a,bc} \nonumber \\
\delta_0 \Phi_{\mu\nu}{}^a = D_{[\mu} \xi_{\nu]}{}^a +
\eta_{\mu\nu}{}^a, &\qquad
&\delta_0 \Omega_\mu{}^{abc} = D_\mu \eta^{abc} \\
\delta_0 h_\mu{}^a = D_\mu \zeta^a + \chi_\mu{}^a, &\qquad
&\delta_0 \omega_\mu{}^{ab} = D_\mu \chi^{ab} \nonumber \\
\delta_0 \Phi_{\mu\nu} = D_{[\mu} \xi_{\nu]}, &\qquad
&\delta_0 h_\mu = D_\mu \zeta \nonumber
\end{eqnarray}
where all derivatives are now $(A)dS$ covariant ones. As usual, due to
non-commutativity of covariant derivatives this Lagrangian is not
invariant under the initial gauge transformations:
\begin{eqnarray*}
\delta_0 {\cal L}_0 &=& - 3 \kappa (d-2) 
\left\{ \phantom{|}^{\mu\nu}_{ab} \right\} ( 2 \Omega_\mu{}^{abc,d}
\xi_\nu{}^{cd} - \eta^{abc,d} \Phi_{\mu\nu}{}^{cd} ) + 3 \kappa (d-1)
( \omega_\mu{}^{\mu,ab} \zeta^{ab} - \chi^{\mu,ab} h_\mu{}^{ab} )
\\
&& \kappa \left\{ \phantom{|}^{\mu\nu}_{ab} \right\} [ 3 (d-3) ( 2
\Omega_\mu{}^{abc} \xi_\nu{}^c + \eta^{abc} \Phi_{\mu\nu}{}^c ) -
(d-2) ( \omega_\mu{}^{ab} \zeta_\nu - \chi^{ab} h_{\mu\nu}) ]
\end{eqnarray*}
but we will take this non-invariance into account later on.

To construct gauge invariant description of massive particles we
proceed by adding cross terms of order $m$ (i.e. terms with the
coefficients with dimension of mass) to the Lagrangian. As we have
already noted, one has to introduce such cross terms for the nearest
neighbours only, i.e main gauge field with primary ones, primary with
secondary and so on. To simplify the presentation we consider these
terms step by step.

$\Phi_{\mu\nu}{}^{ab} \Leftrightarrow \Phi_{\mu\nu}{}^a,h_\mu{}^{ab}$.
In this case additional terms to the Lagrangian could be written in
the following form:
\begin{equation}
{\cal L}_1 = \left\{ \phantom{|}^{\mu\nu\alpha}_{abc} \right\} [
a_1 \Omega_\mu{}^{abc,d} \Phi_{\nu\alpha}{}^d +
a_2 \Phi_{\mu\nu}{}^{ad} \Omega_\alpha{}^{bcd} + 
a_3 \Phi_{\mu\nu}{}^{ad} \omega_\alpha{}^{b,cd} ] +
\left\{ \phantom{|}^{\mu\nu}_{ab} \right\} a_4
\Omega_\mu{}^{abc,d} h_\nu{}^{cd}
\end{equation}
As usual, their non-invariance under the initial gauge transformations
could be compensated by appropriate corrections to gauge
transformations:
\begin{eqnarray}
\delta_1 \Phi_{\mu\nu}{}^{ab} &=& - \frac{4 \alpha_2}{d-2} [
e_{[\mu}{}^{(a} \xi_{\nu]}{}^{b)} + \frac{2}{d} g^{ab} \xi_{[\mu,\nu]}
] + \frac{\beta_2}{6(d-2)}  e_{[\mu}{}^{(a}  \zeta_{\nu]}{}^{b)}
\nonumber \\
\delta_1 \Omega_\mu{}^{abc,d} &=& - \frac{\alpha_2}{d} [ 3 e_\mu{}^d
\eta^{abc} + e_\mu{}^{[a} \eta^{bc]d} - \frac{4}{(d-2)}
g^{d[a} \eta^{bc]}{}_\mu) ] + \nonumber \\
 && + \frac{\beta_2}{3(d-3)} [ e_\mu{}^{[a} \chi^{b,c]d} - 
\frac{1}{d-2} g^{d[a} \chi^{b,c]}{}_\mu ]  \\
\delta_1 \Phi_{\mu\nu}{}^a &=& \alpha_2 \xi_{[\mu,\nu]}{}^a, \qquad
\delta_1 \Omega_\mu{}^{abc} = - 4 \alpha_2 \eta^{abc}{}_\mu  \nonumber
\\
\delta h_\mu{}^{ab} &=& \beta_2 \xi_\mu{}^{ab}, \qquad
\delta_1 \omega_\mu{}^{a,bc} = - \frac{\beta_2}{2}
\eta_\mu{}^{a(b,c)} \nonumber
\end{eqnarray}
provided $a_1 = 4 \alpha_2$, $a_2 = 3 \alpha_2$,
$a_3 = a_4 = - \beta_2$.

$\Phi_{\mu\nu}{}^a,h_\mu{}^{ab} \Leftrightarrow \Phi_{\mu\nu},
h_\mu{}^a$. Now additional terms to the Lagrangian have the form:
\begin{eqnarray}
\Delta {\cal L}_1 &=&
\left\{ \phantom{|}^{\mu\nu\alpha}_{abc} \right\} [ a_5
\omega_\mu{}^{ab} \Phi_{\nu\alpha}{}^c + a_6
\Omega_\mu{}^{abc} \Phi_{\nu\alpha} ] + 
\left\{ \phantom{|}^{\mu\nu}_{ab} \right\} [
a_7 \Omega_\mu{}^{abc} h_\nu{}^c + a_8
\Omega^{abc} \Phi_{\mu\nu}{}^c ] + \nonumber \\
 && + \left\{ \phantom{|}^{\mu\nu}_{ab} \right\} [ a_9
\omega_\mu{}^{a,bc} h_\nu{}^c +
 a_{10} h_\mu{}^{ac} \omega_\nu{}^{bc} ]
\end{eqnarray}
To compensate their non-invariance under the initial gauge
transformations we introduce the following corrections to gauge
transformations:
\begin{eqnarray}
\delta_1 \Phi_{\mu\nu}{}^a &=& 
\frac{\beta_1}{12(d-3)} e_{[\mu}{}^a \zeta_{\nu]} - 
\frac{3\alpha_1}{d-3} e_{[\mu}{}^a \xi_{\nu]}, \qquad
\delta_1 \Omega_\mu{}^{abc} = 
\frac{\beta_3}{6(d-3)} e_\mu{}^{[a} \chi^{bc]} \nonumber \\
\delta_1 h_\mu{}^{ab} &=& \frac{\rho_1}{d-1} [
e_\mu^{(a} \zeta^{b)} - \frac{2}{d} g^{ab} \zeta_\mu ] \nonumber \\
\delta_1 \omega_\mu{}^{a,bc} &=& \frac{\rho_1}{d} [ 
\chi^{a(b} e_\mu{}^{c)} + \frac{1}{d-1} ( 2 g^{bc}
\chi_\mu{}^a - g^{a(b} \chi_\mu{}^{c)} )] \\
\delta_1 h_\mu{}^a &=& \beta_1 \xi_\mu{}^a + \rho_1 \zeta_\mu{}^a,
\qquad \delta_1 \omega_\mu{}^{ab} = - \frac{\beta_1}{2}
\eta_\mu{}^{ab} + \rho_1 \chi^{[a,b]}{}_\mu \nonumber \\
\delta_1 \Phi_{\mu\nu} &=& \alpha_1 \xi_{[\mu,\nu]}, \qquad
\delta_1 \Omega^{abc} = - 3 \alpha_1 \eta^{abc} \nonumber
\end{eqnarray}
provided $a_5 = a_7 = \beta_1/2$, 
$a_6 = a_8 = - 3 \alpha_1$, $a_9 = a_{10} = - 2 \rho_1$.

$\Phi_{\mu\nu}, h_\mu{}^a \Leftrightarrow \Phi_{\mu\nu}, h_\mu$.
Finally, we add to the Lagrangian terms (we already familiar with):
\begin{equation}
\Delta {\cal L}_1 = \left\{ \phantom{|}^{\mu\nu}_{ab} \right\} [
a_{11} \omega_\mu{}^{ab} h_\nu + a_{12} \omega^{ab} \Phi_{\mu\nu}  
] + a_{13} \omega^{ab} h_{ab} 
\end{equation}
and corresponding corrections to gauge transformations:
\begin{equation}
\delta h_\mu{}^a = \frac{2\rho_0}{d-2} e_\mu{}^a \zeta, \qquad
\delta h_\mu = \rho_0 \zeta_\mu + \beta_0 \xi_\mu, \qquad
\delta \omega^{ab} = - 2 \rho_0 \chi^{ab}
\end{equation}
where $a_{11} = a_{13} = 2 \rho_0$, $a_{12} = \beta_0/2$.

Collecting all pieces together, we obtain complete set of cross terms:
\begin{eqnarray}
{\cal L}_1 &=& \left\{ \phantom{|}^{\mu\nu\alpha}_{abc} \right\} [
4 \alpha_2 \Omega_\mu{}^{abc,d} \Phi_{\nu\alpha}{}^d + 3 \alpha_2
\Phi_{\mu\nu}{}^{ad} \Omega_\alpha{}^{bcd} - \beta_2
\Phi_{\mu\nu}{}^{ad} \omega_\alpha{}^{b,cd} ] - \beta_2
\left\{ \phantom{|}^{\mu\nu}_{ab} \right\}
\Omega_\mu{}^{abc,d} h_\nu{}^{cd} + \nonumber \\
 && + \left\{ \phantom{|}^{\mu\nu\alpha}_{abc} \right\} [
\frac{\beta_1}{2} \omega_\mu{}^{ab} \Phi_{\nu\alpha}{}^c - 3 \alpha_1
\Omega_\mu{}^{abc} \Phi_{\nu\alpha} ] + 
\left\{ \phantom{|}^{\mu\nu}_{ab} \right\} [ \frac{\beta_1}{2}
\Omega_\mu{}^{abc} h_\nu{}^c - 3 \alpha_1
\Omega^{abc} \Phi_{\mu\nu}{}^c ] + \nonumber \\
 && + \left\{ \phantom{|}^{\mu\nu}_{ab} \right\} [ - 2 \rho_1
\omega_\mu{}^{a,bc} h_\nu{}^c - 2 \rho_1 h_\mu{}^{ac}
\omega_\nu{}^{bc} + 2 \rho_0 \omega_\mu{}^{ab} h_\nu + 
\frac{\beta_0}{2} \omega^{ab} \Phi_{\mu\nu}  ] + 2 \rho_0 \omega^{ab}
h_{ab} 
\end{eqnarray}
Now, as we have achieved cancellation of all variations of order $m$
$\delta_0 {\cal L}_1 + \delta_1 {\cal L}_0$, we have to take care on
variations of order $m^2$ (including contribution from kinetic terms
due to non-commutativity of covariant derivatives) $\delta_0 {\cal
L}_0 + \delta_1 {\cal L}_1$. As in the previous case, there are no any
explicit mass-like terms allowed here, but complete invariance of the
Lagrangian could be achieved just by adjusting the values of remaining
free parameters $\alpha_{1,2}$, $\beta_{0,1,2}$ and $\rho_{0,1}$:
$$
\beta_1 = - 2 \sqrt{\frac{d-1}{d-2}} \beta_2, \quad 
\beta_0 = - \sqrt{\frac{6(d-1)}{d-3}} \beta_2, \quad
\rho_1 = 2 \sqrt{\frac{d-1}{d-2}} \alpha_2, \quad 
\rho_0 = \sqrt{\frac{3(d-2)}{2(d-3)}} \alpha_1
$$
$$
24 \alpha_2{}^2 - \beta_2{}^2 = 6 (d-2) \kappa, \qquad
12 (d+1) \alpha_2{}^2 - 3 d \alpha_1{}^2 = d (d+1) \kappa
$$
The role that each of the parameters plays could be easily seen from
the Figure 2.
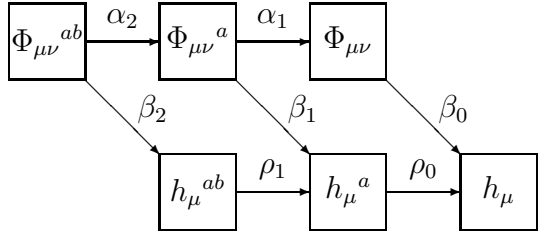
\begin{figure}[htb]
\begin{center}
\begin{picture}(90,32)
\put(10,21){\framebox(10,10)[]{$\Phi_{\mu\nu}{}^{ab}$}}
\put(30,21){\framebox(10,10)[]{$\Phi_{\mu\nu}{}^a$}}
\put(50,21){\framebox(10,10)[]{$\Phi_{\mu\nu}$}}
\put(30,1){\framebox(10,10)[]{$h_\mu{}^{ab}$}}
\put(50,1){\framebox(10,10)[]{$h_\mu{}^a$}}
\put(70,1){\framebox(10,10)[]{$h_\mu$}}
\put(20,26){\vector(1,0){10}}
\put(22,26){\makebox(6,6)[]{$\alpha_2$}}
\put(20,21){\vector(1,-1){10}}
\put(26,14){\makebox(6,6)[]{$\beta_2$}}
\put(40,26){\vector(1,0){10}}
\put(42,26){\makebox(6,6)[]{$\alpha_1$}}
\put(40,21){\vector(1,-1){10}}
\put(46,14){\makebox(6,6)[]{$\beta_1$}}
\put(40,6){\vector(1,0){10}}
\put(42,6){\makebox(6,6)[]{$\rho_1$}}
\put(60,21){\vector(1,-1){10}}
\put(66,14){\makebox(6,6)[]{$\beta_0$}}
\put(60,6){\vector(1,0){10}}
\put(62,6){\makebox(6,6)[]{$\rho_0$}}
\end{picture}
\end{center}
\caption{General massive theory for $Y(3,1)$ tensor}
\end{figure}
Now we are ready to analyze the results obtained. First of all, note
that the massless limit (i.e decoupling of main gauge fields from all
others) requires $\alpha_2 = \beta_2 = 0$. As the first of last two
relations clearly shows this is possible in flat Minkowski space
($\kappa = 0$) only. In this, for non-zero values of cosmological
constant there exists a number of partially massless limits.

In $dS$ space ($\kappa >0$) one can put $\beta_2 = 0$ (and this
simultaneously gives $\beta_1 = \beta_0 = 0$). In this complete system
decompose into two disconnected subsystems, as shown on the Figure 3.
\begin{figure}[htb]
\begin{center}
\begin{picture}(90,32)
\put(10,21){\framebox(10,10)[]{$\Phi_{\mu\nu}{}^{ab}$}}
\put(30,21){\framebox(10,10)[]{$\Phi_{\mu\nu}{}^a$}}
\put(50,21){\framebox(10,10)[]{$\Phi_{\mu\nu}$}}
\put(30,1){\framebox(10,10)[]{$h_\mu{}^{ab}$}}
\put(50,1){\framebox(10,10)[]{$h_\mu{}^a$}}
\put(70,1){\framebox(10,10)[]{$h_\mu$}}
\put(20,26){\vector(1,0){10}}
\put(22,26){\makebox(6,6)[]{$\alpha_2$}}
\put(40,26){\vector(1,0){10}}
\put(42,26){\makebox(6,6)[]{$\alpha_1$}}
\put(40,6){\vector(1,0){10}}
\put(42,6){\makebox(6,6)[]{$\rho_1$}}
\put(60,6){\vector(1,0){10}}
\put(62,6){\makebox(6,6)[]{$\rho_0$}}
\end{picture}
\end{center}
\caption{Partially massless limit in $dS$ space}
\end{figure}
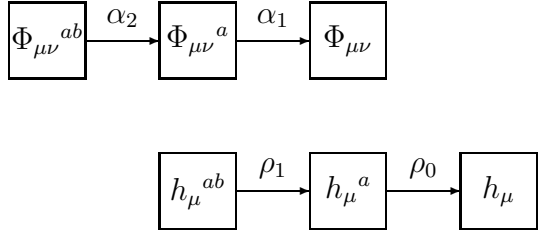
One of them, with the fields $\Phi_{\mu\nu}{}^{ab}$, 
$\Phi_{\mu\nu}{}^a$ and $\Phi_{\mu\nu}$ gives new example of partially
massless theory with the Lagrangian:
\begin{eqnarray}
{\cal L} &=& {\cal L}_0 (\Phi_{\mu\nu}{}^{ab}) + {\cal L}_0
(\Phi_{\mu\nu}{}^a) + {\cal L}_0 (\Phi_{\mu\nu}) + \nonumber \\
 && + \alpha_2 \left\{ \phantom{|}^{\mu\nu\alpha}_{abc} \right\} [
4 \Omega_\mu{}^{abc,d} \Phi_{\nu\alpha}{}^d + 3 
\Phi_{\mu\nu}{}^{ad} \Omega_\alpha{}^{bcd} ]  - \nonumber \\
 && - 3 \alpha_1 [ \left\{ \phantom{|}^{\mu\nu\alpha}_{abc} \right\}
\Omega_\mu{}^{abc} \Phi_{\nu\alpha} + 
\left\{ \phantom{|}^{\mu\nu}_{ab} \right\} \Omega^{abc} 
\Phi_{\mu\nu}{}^c ] 
\end{eqnarray}
where $4\alpha_2{}^2 = (d-2) \kappa$, $3d \alpha_1{}^2 = 2 (d+1) (d-3)
\kappa$, which is invariant under the following gauge transformations
(for simplicity we reproduce here gauge transformations for physical
fields only):
\begin{eqnarray}
\delta \Phi_{\mu\nu}{}^{ab} &=& D_{[\mu} \xi_{\nu]}{}^{ab} +
\eta_{\mu\nu}{}^{(a,b)} - \frac{4 \alpha_2}{d-2} 
e_{[\mu}{}^{(a} \xi_{\nu]}{}^{b)}  \nonumber \\
\delta \Phi_{\mu\nu}{}^a &=& D_{[\mu} \xi_{\nu]}{}^a + 
\eta_{\mu\nu}{}^a + \alpha_2 \xi_{[\mu,\nu]}{}^a  - 
\frac{3\alpha_1}{d-3} e_{[\mu}{}^a \xi_{\nu]}  \\
\delta \Phi_{\mu\nu} &=& D_{[\mu} \xi_{\nu]} + \alpha_1 
\xi_{[\mu,\nu]}  \nonumber
\end{eqnarray}
At the same time, three other fields $h_\mu{}^{ab}$, $h_\mu{}^a$ and
$h_\mu$ give gauge invariant description of partially massless spin 3
particle \cite{SV06,Zin08b}.

One more example of partially massless theory appears if
one put $\alpha_1 = 0$ (and hence $\rho_0 = 0$). In this, complete
system also decompose into two disconnected subsystems as shown on the
Figure 4.
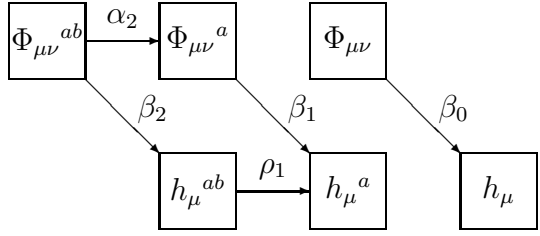
\begin{figure}[htb]
\begin{center}
\begin{picture}(90,32)
\put(10,21){\framebox(10,10)[]{$\Phi_{\mu\nu}{}^{ab}$}}
\put(30,21){\framebox(10,10)[]{$\Phi_{\mu\nu}{}^a$}}
\put(50,21){\framebox(10,10)[]{$\Phi_{\mu\nu}$}}
\put(30,1){\framebox(10,10)[]{$h_\mu{}^{ab}$}}
\put(50,1){\framebox(10,10)[]{$h_\mu{}^a$}}
\put(70,1){\framebox(10,10)[]{$h_\mu$}}
\put(20,26){\vector(1,0){10}}
\put(22,26){\makebox(6,6)[]{$\alpha_2$}}
\put(20,21){\vector(1,-1){10}}
\put(26,14){\makebox(6,6)[]{$\beta_2$}}
\put(40,21){\vector(1,-1){10}}
\put(46,14){\makebox(6,6)[]{$\beta_1$}}
\put(40,6){\vector(1,0){10}}
\put(42,6){\makebox(6,6)[]{$\rho_1$}}
\put(60,21){\vector(1,-1){10}}
\put(66,14){\makebox(6,6)[]{$\beta_0$}}
\end{picture}
\end{center}
\caption{Non-unitary partially massless theory}
\end{figure}

Note, however that in this case we obtain $12 \alpha_2{}^2 = d
\kappa$,  $\beta_2{}^2 = - 4 (d-3) \kappa$, so that such theory (as is
often to be the case) "lives" inside unitary forbidden region.

From the other hand, in $AdS$ space ($\kappa < 0$) one can put
$\alpha_2 = 0$ (and hence $\rho_1 = 0$). In this, decomposition into
two subsystems looks as follows (Figure 5):
\begin{figure}[htb]
\begin{center}
\begin{picture}(90,32)
\put(10,21){\framebox(10,10)[]{$\Phi_{\mu\nu}{}^{ab}$}}
\put(30,21){\framebox(10,10)[]{$\Phi_{\mu\nu}{}^a$}}
\put(50,21){\framebox(10,10)[]{$\Phi_{\mu\nu}$}}
\put(30,1){\framebox(10,10)[]{$h_\mu{}^{ab}$}}
\put(50,1){\framebox(10,10)[]{$h_\mu{}^a$}}
\put(70,1){\framebox(10,10)[]{$h_\mu$}}
\put(20,21){\vector(1,-1){10}}
\put(26,14){\makebox(6,6)[]{$\beta_2$}}
\put(40,26){\vector(1,0){10}}
\put(42,26){\makebox(6,6)[]{$\alpha_1$}}
\put(40,21){\vector(1,-1){10}}
\put(46,14){\makebox(6,6)[]{$\beta_1$}}
\put(60,21){\vector(1,-1){10}}
\put(66,14){\makebox(6,6)[]{$\beta_0$}}
\put(60,6){\vector(1,0){10}}
\put(62,6){\makebox(6,6)[]{$\rho_0$}}
\end{picture}
\end{center}
\caption{Partially massless limit in $AdS$ space}
\end{figure}
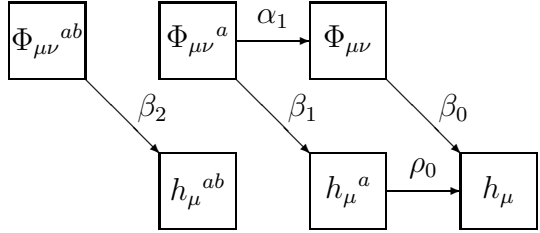

Thus we obtain one more example of partially massless theory with two
fields $\Phi_{\mu\nu}{}^{ab}$ and $h_\mu{}^{ab}$. The Lagrangian:
\begin{equation}
{\cal L} = {\cal L}_0 (\Phi_{\mu\nu}{}^{ab}) + {\cal L}_0
(h_\mu{}^{ab}) - \beta_2 
\left\{ \phantom{|}^{\mu\nu\alpha}_{abc} \right\}
\Phi_{\mu\nu}{}^{ad} \omega_\alpha{}^{b,cd} - \beta_2
\left\{ \phantom{|}^{\mu\nu}_{ab} \right\}
\Omega_\mu{}^{abc,d} h_\nu{}^{cd}
\end{equation}
where $\beta_2{}^2 = - 6 (d-2) \kappa$, is invariant under the
following gauge transformations:
\begin{equation}
\delta \Phi_{\mu\nu}{}^{ab} = D_{[\mu} \xi_{\nu]}{}^{ab} +
\eta_{\mu\nu}{}^{(a,b)}  + \frac{\beta_2}{6(d-2)}  e_{[\mu}{}^{(a} 
\zeta_{\nu]}{}^{b)}, \qquad
\delta h_\mu{}^{ab} = D_\mu \zeta^{ab} + \chi_\mu{}^{ab} + \beta_2
\xi_\mu{}^{ab}
\end{equation}
In this, four remaining fields $\Phi_{\mu\nu}{}^a$, $h_\mu{}^a$,
$\Phi_{\mu\nu}$ and $h_\mu$ just gives the same gauge invariant
massive theory as in the previous section.

\section{Tensor $Y(k,1)$}

For the description of massless particles we introduce main physical
field --- two-form $\Phi_{\mu\nu}{}^{a_1 \dots a_{k-1}} =
\Phi_{\mu\nu}{}^{(k-1)}$ (here and in what follows we will use the
same
condensed notations for tensor objects as in \cite{Zin08b}) which is
completely symmetric and traceless on local indices and auxiliary 
one-form $\Omega_\mu{}^{abc,(k-2)}$ which is completely antisymmetric
on $abc$, traceless on all local indices and satisfies a constraint
$\Omega_\mu{}^{[abc,d](k-3)} = 0$. To have correct number of physical
degrees of freedom massless Lagrangian has to be invariant under the
following gauge transformations:
\begin{equation}
\delta \Phi_{\mu\nu}{}^{(k-1)} = \partial_{[\mu} \xi_{\nu]}{}^{(k-1)}
+ \eta_{\mu\nu}{}^{(1,k-2)}, \qquad \delta \Omega_\mu{}^{abc,(k-2)} =
\partial_\mu \eta^{abc,(k-2)}
\end{equation}
where properties of parameters $\xi$ and $\eta$ correspond to that of
$\Phi_{\mu\nu}$ and $\Omega_\mu$. To find appropriate Lagrangian, we
introduce a tensor $ T_{\mu\nu\alpha}{}^{(k-1)} = \partial_{[\mu}
\Phi_{\nu\alpha]}{}^{(k-1)} $, which is invariant under
$\xi$-transformations, consider an expression 
$ \left\{ \phantom{|}^{\mu\nu\alpha\beta}_{abcd} \right\}
\Omega_\mu{}^{abc,(k-2)} T_{\nu\alpha\beta}{}^{d(k-2)}$ and make a
substitution $T_{\mu\nu\alpha}{}^{(k-1)} \rightarrow 
\Omega_{[\mu,\nu\alpha]}{}^{(1,k-2)}$. We obtain:
$$
\left\{ \phantom{|}^{\mu\nu\alpha\beta}_{abcd} \right\}
\Omega_\mu{}^{abc,(k-2)} T_{\nu\alpha\beta}{}^{d (k-2)} \quad
\Rightarrow \quad
\left\{ \phantom{|}^{\mu\nu\alpha\beta}_{abcd} \right\}
\Omega_\mu{}^{abc,(k-2)} \Omega_{\nu,\alpha\beta}{}^{(d,k-2)} \quad
\Rightarrow
$$
$$
\left\{ \phantom{|}^{\mu\nu}_{ab} \right\}  [ 3
\Omega_\mu{}^{acd(k-2)} \Omega_\nu{}^{bcd,(k-2)} + (k-2)
\Omega_\mu{}^{cde,a(k-3)} \Omega_\nu{}^{cde,b(k-3)} ]
$$
Thus we will look for massless Lagrangian in the form:
\begin{eqnarray*}
{\cal L}_0 &=& a_1 \left\{ \phantom{|}^{\mu\nu}_{ab} \right\}  [ 3
\Omega_\mu{}^{acd,(k-2)} \Omega_\nu{}^{bcd,(k-2)} + (k-2)
\Omega_\mu{}^{cde,a(k-3)} \Omega_\nu{}^{cde,b(k-3)} ] + \\
 &&  + a_2 \left\{ \phantom{|}^{\mu\nu\alpha\beta}_{abcd} \right\}
\Omega_\mu{}^{abc,(k-2)} T_{\nu\alpha\beta}{}^{d(k-2)}
\end{eqnarray*}
It is by construction invariant under the $\xi$-transformations, while
invariance under the $\eta$-transformations requires $a_1 = - 3
a_2$. We choose $a_1 = (-1)^{k-1}$, $a_2 = - (-1)^{k-1}/3$ and obtain
finally:
\begin{eqnarray}
(-1)^{k-1} {\cal L}_0 &=& \left\{ \phantom{|}^{\mu\nu}_{ab} \right\} 
[ 3 \Omega_\mu{}^{acd,(k-2)} \Omega_\nu{}^{bcd,(k-2)} + (k-2)
\Omega_\mu{}^{cde,a(k-3)} \Omega_\nu{}^{cde,b(k-3)} ] - \nonumber \\
 &&  - \left\{ \phantom{|}^{\mu\nu\alpha\beta}_{abcd} \right\}
\Omega_\mu{}^{abc,(k-2)} \partial_\nu \Phi_{\alpha\beta}{}^{d(k-2)}
\end{eqnarray}

As in the previous cases, it is not possible to deform this massless
theory into $(A)dS$ space without introduction of additional fields,
so we will turn to the general case --- massive particle in $(A)dS$
space with arbitrary cosmological constant. Our first task --- to
determine the set of additional fields which are necessary for gauge
invariant description of such massive particle. Our main gauge field
$Y(k,1)$ has two gauge transformations with parameters $Y(k-1,1)$ and
$Y(k,0)$ with the reducibility $Y(k-1,0)$, thus we need two primary
Goldstone fields $Y(k-1,1)$ and $Y(k,0)$. The first one has two own
gauge transformations with parameters $Y(k-2,1)$ and $Y(k-1,0)$ with
reducibility $Y(k-2,0)$, while the second one has one gauge
transformation with parameter $Y(k-1,0)$ only. So we need two
secondary fields $Y(k-2,1)$ and $Y(k-1,0)$ and so on. As in the
previous cases, this procedure stops at vector field $Y(1,0)$, thus we
totally have to introduce fields $Y(l,1)$ and $Y(l,0)$ with $1 \le l
\le k$. 

Let us start with the sum of kinetic terms for all these fields:
\begin{eqnarray}
{\cal L}_0 &=& \sum_{l=2}^{k} {\cal L}_0 (\Phi_{\mu\nu}{}^{(l-1)})
- \Omega_{abc}{}^2 + \left\{ \phantom{|}^{\mu\nu\alpha}_{abc}
\right\} \Omega^{abc} D_\mu \Phi_{\nu\alpha} + \nonumber \\
 && + \sum_{l=2}^{k} {\cal L}_0 (h_\mu{}^{(l-1)})
 + \frac{1}{2} \omega_{ab}{}^2 - \left\{ \phantom{|}^{\mu\nu}_{ab}
\right\} \omega^{ab} D_\mu h_\nu
\end{eqnarray}
\begin{eqnarray*}
(-1)^l {\cal L}_0 (\Phi_{\mu\nu}{}^{(l)}) &=& 
\left\{ \phantom{|}^{\mu\nu}_{ab} \right\}  [ 3
\Omega_\mu{}^{acd,(l-1)} \Omega_\nu{}^{bcd,(l-1)} + (l-1)
\Omega_\mu{}^{cde,a(l-2)} \Omega_\nu{}^{cde,b(l-2)} ] - \\
 &&  -  \left\{ \phantom{|}^{\mu\nu\alpha\beta}_{abcd} \right\}
\Omega_\mu{}^{abc,(l-1)} D_\nu \Phi_{\alpha\beta}{}^{d(l-1)}  \\
(-1)^l {\cal L}_0 (h_\mu{}^{(l)}) &=& - 
\left\{ \phantom{|}^{\mu\nu}_{ab} \right\}  [
\omega_\mu{}^{c,a(l-1)} \omega_\nu{}^{c,b(l-1)} + \frac{1}{l}
\omega_\mu{}^{a,(l)} \omega_\nu{}^{b,(l)} ] + \\
 && + 2 \left\{ \phantom{|}^{\mu\nu\alpha}_{abc} \right\}
\omega_\mu{}^{a,b(l-1)} D_\nu h_\alpha{}^{c(l-1)}
\end{eqnarray*}
as well as appropriate set of initial gauge transformations:
\begin{eqnarray}
\delta \Phi_{\mu\nu}{}^{(l)} &=& D_{[\mu} \xi_{\nu]}{}^{(l)} +
\eta_{\mu\nu}{}^{(1,l-1)}, \qquad \delta \Omega_\mu{}^{abc,(l-1)} =
D_\mu \eta^{abc,(l-1)}, \qquad 
\delta \Phi_{\mu\nu} = D_{[\mu} \xi_{\nu]} \nonumber \\
\delta h_\mu{}^{(l)} &=& D_\mu \zeta^{(l)} + \chi_\mu{}^{(l)}, \qquad
\delta \omega_\mu{}^{a,(l)} = D_\mu \chi^{a,(l)}, \qquad
\delta h_\mu = D_\mu \zeta
\end{eqnarray}
where all derivatives are now $(A)dS$ covariant ones. Due to
non-commutativity of covariant derivatives this Lagrangian is not
invariant under the initial gauge transformations:
\begin{eqnarray*}
\delta_0 {\cal L}_0 &=& \sum_{l=2}^{k} (-1)^l \kappa 
\left\{ \phantom{|}^{\mu\nu}_{ab} \right\}[ 3 (d+l-4) ( -2
\Omega_\mu{}^{abc,(l-1)} \xi_\nu{}^{c(l-1)} + \eta^{abc,(l-1)}
\Phi_{\mu\nu}{}^{c(l-1)} ) + \nonumber \\
 && \qquad \qquad \qquad +  2 (d+l-3) ( \omega_\mu{}^{a,b(l-1)}
\zeta_\nu{}^{(l-1)} - \frac{l+1}{l} \chi^{\mu,(l)} h_\mu{}^{(l)} ) ]
\end{eqnarray*}
but we will take this non-invariance into account later on.\

To proceed with the construction of gauge invariant description of
massive particle, we have to add to the Lagrangian cross terms of
order $m$ (i.e. with the coefficients having dimension of mass). As we
have already noted above, one has to introduce such cross terms for
the nearest neighbours only (i.e. main field with primary ones,
primary with secondary and so on). For the case at hands, this means
introduction cross terms between pairs  $Y(l+1,1)$, $Y(l+2,0)$ and
$Y(l,1)$, $Y(l+1,0)$. Moreover, due to symmetry and tracelessness
properties of the fields, there exists such terms for three possible
cases $Y(l+1,1) \Leftrightarrow Y(l,1)$, 
$Y(l+1,1) \Leftrightarrow Y(l+1,0)$,
$Y(l+2,0) \Leftrightarrow Y(l+1,0)$ only. We consider these three
possibilities in turn.

$\Omega_\mu{}^{abc,(l-1)}, \Phi_{\mu\nu}{}^{(l)} \Leftrightarrow
\Omega_\mu{}^{abc,(l-2)}, \Phi_{\mu\nu}{}^{(l-1)}$. Here additional
terms to the Lagrangian could be written as follows:
\begin{equation}
{\cal L}_1 = (-1)^l \left\{ \phantom{|}^{\mu\nu\alpha}_{abc} \right\}
[ a_{1l} \Omega_\mu{}^{abc,(l-1)} \Phi_{\nu\alpha}{}^{(l-1)} + a_{2l}
\Omega_\mu{}^{abd,(l-2)} \Phi_{\nu\alpha}{}^{cd(l-2)} ]
\end{equation}
Their non-invariance under the initial gauge transformations could be
compensated by the following corrections to gauge transformations:
\begin{eqnarray}
\delta_1 \Phi_{\mu\nu}{}^{(l)} &=& - 
\frac{(l+2)\alpha_l}{(l-1)(d+l-4)} [
e_{[\mu}{}^{(1} \xi_{\nu]}{}^{l-1)} + \frac{2}{d+2l-4} 
\xi_{[\mu,\nu]}{}^{(l-2} g^{12)} ] \nonumber \\
\delta_1 \Omega_\mu{}^{abc,(l-1)} &=& - \frac{\alpha_l}{(l-1)d} [
3 \eta^{abc,(l-2} e_\mu^{1)} + e_\mu{}^{[a} \eta^{bc](1,l-2)} - Tr] \\
\delta_1 \Phi_{\mu\nu}{}^{(l-1)} &=& \alpha_l 
\xi_{[\mu,\nu]}{}^{(l-1)}, \qquad 
\delta_1 \Omega_\mu{}^{abc,(l-2)} = - \frac{(l+2)\alpha_l}{l-1}
\eta^{abc,(l-2)}{}_\mu \nonumber
\end{eqnarray}
provided:
$$
a_{1l} = \frac{(l+2)}{(l-1)} \alpha_l, \qquad a_{2l} = 3 \alpha_l
$$

$\Omega_\mu{}^{abc,(l-1)}, \Phi_{\mu\nu}{}^{(l)} \Leftrightarrow
\omega_\mu{}^{a,(l)}, h_\mu{}^{(l)}$. This time additional terms to
the Lagrangian have the form:
\begin{equation}
{\cal L}_1 = 
(-1)^l [ a_{3l} \left\{ \phantom{|}^{\mu\nu}_{ab} \right\} 
\Omega_\mu{}^{abc,(l-1)} h_\nu{}^{c(l-1)} + a_{4l}
\left\{ \phantom{|}^{\mu\nu\alpha}_{abc} \right\}
\omega_\mu{}^{a,b(l-1)} \Phi_{\nu\alpha}{}^{c(l-1)} ]
\end{equation}
and their non-invariance under the initial gauge transformations could
be compensated by:
\begin{eqnarray}
\delta_1 \Phi_{\mu\nu}{}^{(l)} &=& \frac{\beta_l}{6(d+l-4)} 
e_{[\mu}{}^{(1} \zeta_{\nu]}{}^{l-1)} \nonumber \\
\delta_1 \Omega_\mu{}^{abc,(l-1)} &=& \frac{\beta_l}{3(d-3)} [
e_\mu{}^{[a} \chi^{b,c](l-1)} - Tr ] \\
\delta_1 h_\mu{}^{(l)} &=& \beta_l \xi_\mu{}^{(l)}, \qquad
\delta_1 \omega_\mu{}^{a,(l)} = - \frac{\beta_l}{2}
\eta_\mu{}^{a(1,l-1)} \nonumber
\end{eqnarray}
provided $ a_{3l} = a_{4l} = - \beta_l$.

$\omega_\mu{}^{a,(l+1)}, h_\mu{}^{(l+1)} \Leftrightarrow
\omega_\mu{}^{a,(l)}, h_\mu{}^{(l)}$. This case (that has already been
considered in \cite{Zin08b}) requires additional terms to the
Lagrangian in the form:
\begin{equation}
{\cal L}_1 = (-1)^l
\left\{ \phantom{|}^{\mu\nu}_{ab} \right\}  [ a_{5l}
\omega_\mu{}^{a,b(l)} h_\nu{}^{(l)} + a_{6l}
\omega_\mu{}^{a,(l)} h_\nu{}^{b(l)} ]
\end{equation}
as well as the following corrections to gauge transformations:
\begin{eqnarray}
\delta_1 h_\mu{}^{(l+1)} &=& \frac{(l+1)\rho_l}{l(d+l-2)} [
e_\mu{}^{(1} \xi^{l)} - Tr], \qquad \delta_1 \omega_\mu{}^{a,(l+1)} = 
\frac{(l+1)\rho_l}{l(d+l-1)} [ \eta^{a,(l} e_\mu{}^{1)} - Tr ]
\nonumber \\
\delta_1 h_\mu{}^{(l)} &=& \rho_l \xi_\mu{}^{(l)}, \qquad
\delta_1 \omega_\mu{}^{a,(l)} = \frac{\rho_l}{l} [ 
\eta_\mu{}^{a(l)} + (l+1) \eta^{a,(l)}{}_\mu  - Tr]
\end{eqnarray}
where $a_{5l} = a_{6l} = \frac{2(l+1)}{l} \rho_l$.

Collecting all pieces together, we obtain finally:
\begin{eqnarray}
{\cal L}_1 &=& \sum_{l=2}^{k-1} (-1)^l \left[
\left\{ \phantom{|}^{\mu\nu\alpha}_{abc} \right\} \alpha_l [
\frac{l+2}{l-1} \Omega_\mu{}^{abc,(l-1)} \Phi_{\nu\alpha}{}^{(l-1)} +
3 \Omega_\mu{}^{abd,(l-2)} \Phi_{\nu\alpha}{}^{cd(l-2)} ] - \right.
\nonumber \\
 && \qquad \qquad - \beta_l [ \left\{ \phantom{|}^{\mu\nu}_{ab}
\right\}  \Omega_\mu{}^{abc,(l-1)} h_\nu{}^{c(l-1)} + 
\left\{ \phantom{|}^{\mu\nu\alpha}_{abc} \right\}
\omega_\mu{}^{a,b(l-1)} \Phi_{\nu\alpha}{}^{c(l-1)} ] + \nonumber \\
 && \qquad \qquad \left. +  \frac{2(l+1)}{l} \rho_l \left\{
\phantom{|}^{\mu\nu}_{ab} \right\} [ \omega_\mu{}^{a,b(l)}
h_\nu{}^{(l)} + \omega_\mu{}^{a,(l)} h_\nu{}^{b(l)} ] \right] - \\
 && - \left\{ \phantom{|}^{\mu\nu\alpha}_{abc} \right\} 3 \alpha_1
\Omega_\mu{}^{abc} \Phi_{\nu\alpha} + 
\left\{ \phantom{|}^{\mu\nu}_{ab} \right\} [ - 3 \alpha_1
\Omega^{abc} \Phi_{\mu\nu}{}^c +  2 \rho_0
\omega_\mu{}^{ab} h_\nu + \beta_0 \omega^{ab} \Phi_{\mu\nu}  
] + 2 \rho_0 \omega^{ab} h_{ab} \nonumber
\end{eqnarray}
As for the corrections to gauge transformations, we once again
restrict ourselves with the transformations for physical fields
only:
\begin{eqnarray}
\delta_1 \Phi_{\mu\nu}{}^{(l)} &=& \alpha_{l+1} 
\xi_{[\mu,\nu]}{}^{(l)}  - \frac{(l+2)\alpha_l}{(l-1)(d+l-4)} [
e_{[\mu}{}^{(1} \xi_{\nu]}{}^{l-1)} - Tr ] + 
\frac{\beta_l}{6(d+l-4)} e_{[\mu}{}^{(1} \zeta_{\nu]}{}^{l-1)}
\nonumber \\
\delta_1 \Phi_{\mu\nu}{}^a &=& \alpha_2 \xi_{[\mu,\nu]}{}^a -
\frac{3\alpha_1}{d-3} e_{[\mu}{}^a \xi_{\nu]} + \frac{\beta_1}{6(d-3)}
e_{[\mu}{}^a \zeta_{\nu]}, \qquad 
\delta_1 \Phi_{\mu\nu} = \alpha_1 \xi_{[\mu,\nu]} \nonumber \\
\delta_1 h_\mu{}^{(l)} &=& \beta_l \xi_\mu{}^{(l)} + \rho_l
\zeta_\mu{}^{(l)} + \frac{l\rho_{l-1}}{(l-1)(d+l-3)} [ e_\mu{}^{(1}
\zeta^{l-1)} - Tr]  \\
\delta_1 h_\mu{}^a &=& \beta_1 \xi_\mu{}^a + \rho_1 \zeta_\mu{}^a +
\frac{\rho_0}{d-2} e_\mu{}^a \zeta, \qquad
\delta_1 h_\mu = \beta_0 \xi_\mu + \rho_0 \zeta_\mu  \nonumber
\end{eqnarray}

Now, having achieved cancellation of all variations of order $m$
$\delta_0 {\cal L}_1 + \delta_1 {\cal L}_0 = 0$, we have to take care
on variations of order $m^2$ (including contribution of kinetic terms
due to non-commutativity of covariant derivatives) $\delta_0 {\cal
L}_0 + \delta_1 {\cal L}_1$. As in the previous cases, complete
invariance of the Lagrangian could be achieved without introduction of
any explicit mass-like terms into the Lagrangian (and appropriate
corrections to gauge transformations). Indeed, rather long
calculations give four relations:
$$
\alpha_l \beta_{l-1} = - \beta_l \rho_{l-1}
$$
$$
(l+1)(d+l-3) \beta_l \rho_l = -
(l+3)(d+l-2) \alpha_{l+1} \beta_{l+1}
$$
$$
- \frac{6(l+3)(d+l-4)(d+2l)}{l(d+l-3)(d+2l-2)} \alpha_{l+1}{}^2 +
\frac{6(l+2)}{l-1} \alpha_l{}^2 - \beta_l{}^2 = 6 \kappa (d+l-4)
$$
$$
\frac{d+l-3}{3(d+l-4)} \beta_l{}^2 + 
\frac{2(l+1)(d+l-3)(d+2l)}{l(d+l-2)(d+2l-2)} \rho_l{}^2 - 
\frac{2l}{l-1} \rho_{l-1}{}^2 = - 2 \kappa (d+l-3)
$$
To solve these relations we proceed as follows. From the first one we
get:
$$
\rho_l = - \frac{\beta_l}{\beta_{l+1}} \alpha_{l+1}
$$
Putting this relation into the second one, we obtain recurrent
relation on parameters $\beta$:
$$
\beta_l{}^2 = \frac{(l+3)(d+l-2)}{(l+1)(d+l-3)} \beta_{l+1}{}^2
$$
This allows us to express all parameters $\beta_l$ in terms of
$\beta_{k-1}$:
$$
\beta_l{}^2 = \frac{k(k+1)(d+k-4)}{(l+1)(l+2)(d+l-3)} \beta_{k-1}{}^2
$$
In this, one can show that fourth equation is equivalent to third one.
When all parameters $\beta$ are known, the third equation becomes
recurrent relation on parameters $\alpha$ and this allows us
 (taking into account that $\alpha_k = 0$) to express all
$\alpha_l$ in terms of $\alpha_{k-1}$. Let us introduce a notation
$M^2 = \frac{k(k+1)}{k-2} \alpha_{k-1}{}^2$, then the expression for
$\alpha_l$ could be written as follows:
$$
\alpha_l{}^2 = \frac{(l-1)(d+k+l-3)}{(l+1)(l+2)(d+2l-2)} [ M^2 - 
(k-l-1)(d+k+l-4) \kappa ]
$$
Thus we are managed to express all parameters in terms of two main
ones $\beta_{k-1}$ and $M$ (or $\alpha_{k-1}$), in this the following
relation must hold:
$$
6 M^2 - k \beta_{k-1}{}^2 = 6 k (d+k-5) \kappa
$$
Now we are ready to analyze the results obtained. In complete theory
we have three sets of parameters $\alpha$, $\beta$ and $\rho$ and the
roles they play could be easily seen from the Figure 6.
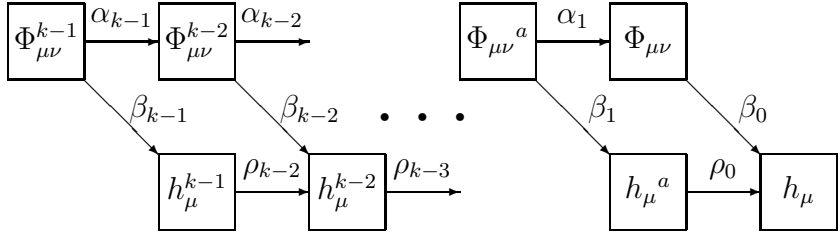
\begin{figure}[htb]
\begin{center}
\begin{picture}(140,32)
\put(10,21){\framebox(10,10)[]{$\Phi_{\mu\nu}^{k-1}$}}
\put(30,21){\framebox(10,10)[]{$\Phi_{\mu\nu}^{k-2}$}}
\put(30,1){\framebox(10,10)[]{$h_\mu^{k-1}$}}
\put(50,1){\framebox(10,10)[]{$h_\mu^{k-2}$}}
\put(20,26){\vector(1,0){10}}
\put(22,26){\makebox(6,6)[]{$\alpha_{k-1}$}}
\put(40,26){\vector(1,0){10}}
\put(42,26){\makebox(6,6)[]{$\alpha_{k-2}$}}
\put(20,21){\vector(1,-1){10}}
\put(27,14){\makebox(6,6)[]{$\beta_{k-1}$}}
\put(40,21){\vector(1,-1){10}}
\put(47,14){\makebox(6,6)[]{$\beta_{k-2}$}}
\put(40,6){\vector(1,0){10}}
\put(42,6){\makebox(6,6)[]{$\rho_{k-2}$}}
\put(60,6){\vector(1,0){10}}
\put(62,6){\makebox(6,6)[]{$\rho_{k-3}$}}
\multiput(60,16)(5,0){3}{\circle*{1}}
\put(70,21){\framebox(10,10)[]{$\Phi_{\mu\nu}{}^a$}}
\put(90,21){\framebox(10,10)[]{$\Phi_{\mu\nu}$}}
\put(90,1){\framebox(10,10)[]{$h_\mu{}^a$}}
\put(110,1){\framebox(10,10)[]{$h_\mu$}}
\put(80,26){\vector(1,0){10}}
\put(82,26){\makebox(6,6)[]{$\alpha_1$}}
\put(80,21){\vector(1,-1){10}}
\put(86,14){\makebox(6,6)[]{$\beta_1$}}
\put(100,21){\vector(1,-1){10}}
\put(106,14){\makebox(6,6)[]{$\beta_0$}}
\put(100,6){\vector(1,0){10}}
\put(102,6){\makebox(6,6)[]{$\rho_0$}}
\end{picture}
\end{center}
\caption{General massive $Y(k,1)$ theory}
\end{figure}

First of all note, that massless limit (that requires $M \rightarrow
0$ and $\beta_{k-1} \rightarrow 0$ simultaneously) is indeed possible
in flat Minkowski space only, while for non-zero values of
cosmological constant we can obtain a number of partially massless
theories. In $AdS$ space ($\kappa < 0$) one can put $\alpha_{k-1} = 0$
(and this gives $\rho_{k-2} = 0$), in this two fields 
$\Phi_{\mu\nu}{}^{(k-1)}$ and $h_\mu{}^{(k-1)}$ decouple and describe
partially massless theory with the Lagrangian (Figure 7):
\begin{equation}
{\cal L} = {\cal L}_0 (\Phi_{\mu\nu}{}^{(k-1)},h_\mu{}^{(k-1)})+
(-1)^{k} \beta_{k-1}
 [ \left\{ \phantom{|}^{\mu\nu}_{ab} \right\} 
 \Omega_\mu{}^{abc,(k-2)} h_\nu{}^{c(k-2)} + 
\left\{ \phantom{|}^{\mu\nu\alpha}_{abc} \right\}
\omega_\mu{}^{a,b(k-2)} \Phi_{\nu\alpha}{}^{c(k-2)} ] 
\end{equation}
which is invariant under the following gauge transformations:
\begin{eqnarray}
\delta \Phi_{\mu\nu}{}^{(k-1)} &=& D_{[\mu} \xi_{\nu]}{}^{(k-1)} +
\eta_{\mu\nu}{}^{(1,k-2)} + 
\frac{\beta_{k-1}}{6(d+k-5)} e_{[\mu}{}^{(1} \zeta_{\nu]}{}^{k-2)}
 \nonumber \\
\delta h_\mu{}^{(k-1)} &=& D_\mu \zeta^{(k-1)} + \chi_\mu{}^{(k-1)}
+ \beta_{k-1} \xi_\mu{}^{(k-1)} 
\end{eqnarray}
At the same time, all other fields just give gauge invariant
description of massive $\Phi_{\mu\nu}{}^{(k-2)}$ tensor.
\begin{figure}[htb]
\begin{center}
\begin{picture}(140,32)
\put(10,21){\framebox(10,10)[]{$\Phi_{\mu\nu}^{k-1}$}}
\put(30,21){\framebox(10,10)[]{$\Phi_{\mu\nu}^{k-2}$}}
\put(30,1){\framebox(10,10)[]{$h_\mu^{k-1}$}}
\put(50,1){\framebox(10,10)[]{$h_\mu^{k-2}$}}
\put(40,26){\vector(1,0){10}}
\put(42,26){\makebox(6,6)[]{$\alpha_{k-2}$}}
\put(20,21){\vector(1,-1){10}}
\put(27,14){\makebox(6,6)[]{$\beta_{k-1}$}}
\put(40,21){\vector(1,-1){10}}
\put(47,14){\makebox(6,6)[]{$\beta_{k-2}$}}
\put(60,6){\vector(1,0){10}}
\put(62,6){\makebox(6,6)[]{$\rho_{k-3}$}}
\multiput(60,16)(5,0){3}{\circle*{1}}
\put(70,21){\framebox(10,10)[]{$\Phi_{\mu\nu}{}^a$}}
\put(90,21){\framebox(10,10)[]{$\Phi_{\mu\nu}$}}
\put(90,1){\framebox(10,10)[]{$h_\mu{}^a$}}
\put(110,1){\framebox(10,10)[]{$h_\mu$}}
\put(80,26){\vector(1,0){10}}
\put(82,26){\makebox(6,6)[]{$\alpha_1$}}
\put(80,21){\vector(1,-1){10}}
\put(86,14){\makebox(6,6)[]{$\beta_1$}}
\put(100,21){\vector(1,-1){10}}
\put(106,14){\makebox(6,6)[]{$\beta_0$}}
\put(100,6){\vector(1,0){10}}
\put(102,6){\makebox(6,6)[]{$\rho_0$}}
\end{picture}
\end{center}
\caption{Partially massless limit in $AdS$ space}
\end{figure}
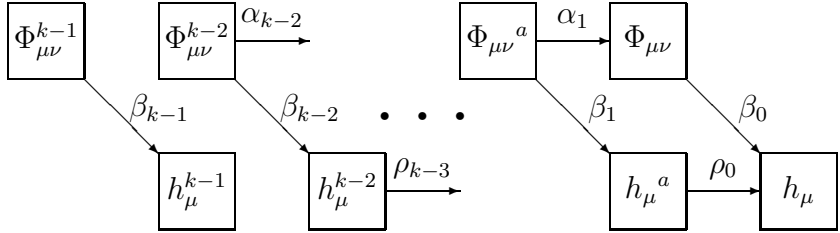

On the other hand, in $dS$ space ($\kappa > 0$) one can put
$\beta_{k-1} = 0$ (and this results in $\beta_l = 0$ for all $l$). In
this case complete system decompose into two disconnected subsystems
(Figure 8).
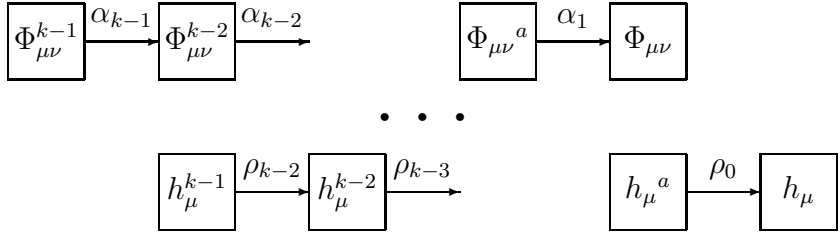
\begin{figure}[htb]
\begin{center}
\begin{picture}(140,32)
\put(10,21){\framebox(10,10)[]{$\Phi_{\mu\nu}^{k-1}$}}
\put(30,21){\framebox(10,10)[]{$\Phi_{\mu\nu}^{k-2}$}}
\put(30,1){\framebox(10,10)[]{$h_\mu^{k-1}$}}
\put(50,1){\framebox(10,10)[]{$h_\mu^{k-2}$}}
\put(20,26){\vector(1,0){10}}
\put(22,26){\makebox(6,6)[]{$\alpha_{k-1}$}}
\put(40,26){\vector(1,0){10}}
\put(42,26){\makebox(6,6)[]{$\alpha_{k-2}$}}
\put(40,6){\vector(1,0){10}}
\put(42,6){\makebox(6,6)[]{$\rho_{k-2}$}}
\put(60,6){\vector(1,0){10}}
\put(62,6){\makebox(6,6)[]{$\rho_{k-3}$}}
\multiput(60,16)(5,0){3}{\circle*{1}}
\put(70,21){\framebox(10,10)[]{$\Phi_{\mu\nu}{}^a$}}
\put(90,21){\framebox(10,10)[]{$\Phi_{\mu\nu}$}}
\put(90,1){\framebox(10,10)[]{$h_\mu{}^a$}}
\put(110,1){\framebox(10,10)[]{$h_\mu$}}
\put(80,26){\vector(1,0){10}}
\put(82,26){\makebox(6,6)[]{$\alpha_1$}}
\put(100,6){\vector(1,0){10}}
\put(102,6){\makebox(6,6)[]{$\rho_0$}}
\end{picture}
\end{center}
\caption{Partially massless limit in $dS$ space}
\end{figure}
One subsystem with the fields $\Phi_{\mu\nu}{}^{(l)}$,
$0 \le l \le k-1$ gives new example of partially massless theory,
while the partially massless theory described by the second subsystem
$h_\mu{}^{(k)}$, $0 \le l \le k-1$ is already known
\cite{SV06,Zin08b}. Besides, a number of (non-unitary) partially
massless theories appears then one put one of the $\alpha_l = 0$ (and
hence $\rho_{l-1} = 0$). In this, complete system also decompose into
two disconnected subsystems. One of them gives partially massless
theory with the fields $\Phi_{\mu\nu}{}^{(n)}$, $h_\mu{}^{(n)}$, $l
\le n \le k-1$ (Figure 9), while the rest of fields just give massive
theory for the $\Phi_{\mu\nu}{}^{(l-1)}$ tensor.
\begin{figure}[htb]
\begin{center}
\begin{picture}(140,32)
\put(10,21){\framebox(10,10)[]{$\Phi_{\mu\nu}^{k-1}$}}
\put(30,21){\framebox(10,10)[]{$\Phi_{\mu\nu}^{k-2}$}}
\put(30,1){\framebox(10,10)[]{$h_\mu^{k-1}$}}
\put(50,1){\framebox(10,10)[]{$h_\mu^{k-2}$}}
\put(20,26){\vector(1,0){10}}
\put(22,26){\makebox(6,6)[]{$\alpha_{k-1}$}}
\put(40,26){\vector(1,0){10}}
\put(42,26){\makebox(6,6)[]{$\alpha_{k-2}$}}
\put(20,21){\vector(1,-1){10}}
\put(27,14){\makebox(6,6)[]{$\beta_{k-1}$}}
\put(40,21){\vector(1,-1){10}}
\put(47,14){\makebox(6,6)[]{$\beta_{k-2}$}}
\put(40,6){\vector(1,0){10}}
\put(42,6){\makebox(6,6)[]{$\rho_{k-2}$}}
\put(60,6){\vector(1,0){10}}
\put(62,6){\makebox(6,6)[]{$\rho_{k-3}$}}
\multiput(60,16)(5,0){3}{\circle*{1}}
\put(70,21){\framebox(10,10)[]{$\Phi_{\mu\nu}^{(l+1)}$}}
\put(90,21){\framebox(10,10)[]{$\Phi_{\mu\nu}^{(l)}$}}
\put(90,1){\framebox(10,10)[]{$h_\mu^{(l+1)}$}}
\put(110,1){\framebox(10,10)[]{$h_\mu^{(l)}$}}
\put(80,26){\vector(1,0){10}}
\put(82,26){\makebox(6,6)[]{$\alpha_{l+1}$}}
\put(80,21){\vector(1,-1){10}}
\put(86,14){\makebox(6,6)[]{$\beta_{l+1}$}}
\put(100,21){\vector(1,-1){10}}
\put(106,14){\makebox(6,6)[]{$\beta_l$}}
\put(100,6){\vector(1,0){10}}
\put(102,6){\makebox(6,6)[]{$\rho_{l}$}}
\end{picture}
\end{center}
\caption{Example of non-unitary partially massless theory}
\end{figure}
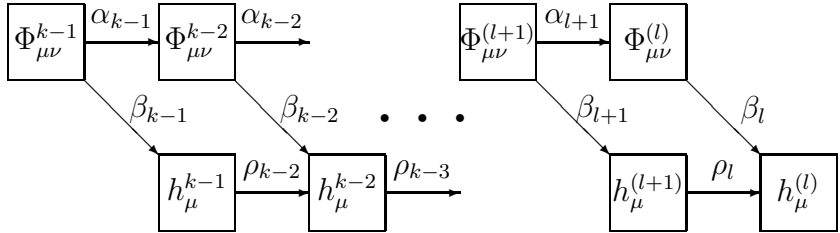

\end{document}